\documentclass{ocg}

\usepackage{tikz}
\usepackage{microtype}
\usepackage{amsmath,amssymb,amstext}
\usepackage{xspace}

\usepackage[utf8]{inputenc}
\usepackage[T1]{fontenc}

\usepackage{paralist}
\usepackage[draft,multiuser,layout=margin]{fixme}

\FXRegisterAuthor{mc}{amc}{MC}
\FXRegisterAuthor{nl}{anl}{NL}
\FXRegisterAuthor{ak}{aak}{AK}

\input{define}

\begin{document}

\title{VALUE AUTOMATA WITH FILTERS$^*$}
\makeatletter
{
  \def\@footnotemark{}
  \let\thefootnote=*
  \d@oldfootnote{Ongoing work.}
  \gdef\d@drtitle{VALUE AUTOMATA WITH FILTERS}
}
\makeatother

\author[e]{Micha\"el Cadilhac}
\author[e]{Andreas Krebs}
\author[z]{Nutan Limaye}

\address[e]{WSI, Universit\"at T\"ubingen
  \email{<\{cadilhac,krebs\}@informatik.uni-tuebingen.de>}}
\address[z]{Indian Institute of Technology Bombay
  \email{<nutan@cse.iitb.ac.in>}}

\thanks[z]{The work of this paper was done during a stay of the third author
  at the Universit\"at T\"ubingen.} 

\maketitle

\begin{abstract}
  We propose to study value automata with filters, a natural generalization
  of regular cost automata to nondeterminism.  Models such as weighted
  automata and Parikh automata appear naturally as specializations.  Results
  on the expressiveness of this model offer a general understanding of the
  behavior of the models that arise as special cases.  A landscape of such
  restrictions is drawn.
\end{abstract}

\section{Introduction}

Through their characteristic functions, formal languages are naturally seen
as giving weights in $\{0, 1\}$ to words.  The transparent correspondence
between these two views is provided by the notion of \emph{weighted
  automata}, a widely studied model firmly rooted in sound algebraic concepts
(see~\cite{sakarovitch09,droste-kuich-vogle09} for recent expositions).
Weighted automata provide an elegant framework to capture some functions from
words to a value set, and their frequent use in the modeling of the
qualitative aspects of real-life
systems~\cite{aminof_reasoning_2010,chatterjee_quantitative_2010}
is a witness of their richness.  In its simpler form, a weighted automaton
can be thought as an automaton where transitions bear integer weights; on
reading a word, the weights are multiplied along the path, and the final
values for all paths labeled by the word are summed.  This model is thus in
some sense ``restricted'' to update its single register by multiplying it by
constants along the path, and summing its different possible values.

In an effort to develop a theory of cost functions with a wider spectrum of
update mechanisms, Alur \emph{et al.}~\cite{alur-reg13} introduced
\emph{regular cost functions}, a highly parametrizable framework in which a
\emph{variant} of weighted automata functions arises as a particular case.
In doing so, they also introduced a \emph{deterministic} model of automata,
\emph{cost register automata} (CRA), and studied in which settings this model
matches regular cost functions.  Similarly to regular cost functions, CRA are
defined with respect to an underlying algebraic structure and restrictions on
the update functions, but in this context, there exists an instance of these
parameters that makes CRA precisely equivalent to weighted automata.

This correspondence between CRA and weighted automata is however not a
straightforward matter of renaming: weighted automata are intrinsically
nondeterministic machines, while CRA are defined as deterministic automata.
With an appropriate generalization of CRA to nondeterminism, in which
weighted automata would arise as an obvious special case, the aforementioned
correspondence would thus express that this case is \emph{determinizable}.

Hence, the starting point of the present research is to propose a
nondeterministic generalization of CRA that encompasses the behavior of
weighted automata in a natural fashion.  By computing on several registers at
once, CRA however provide more information than weighted automata, and simply
aggregating these registers into a single value seems to be a loss.  We thus
propose to add a \emph{filtering} step at the end of the computation: the
vector of registers should lay in a prescribed set.  This allows models such
as reversal-bounded counter machines~\cite{ibarra78}, Parikh
automata~\cite{klaedtke-ruess03}, or finite automata over free
groups~\cite{mitrana-stiebe01} to naturally have a quantitative counterpart.
The fact that the test is made \emph{at the end} of the computation ensures
that the models stay decidable

This framework is sufficiently well-behaved to allow for general results that
echo those to be found within special cases.  It also offers a unified view
of the limits inherent to submodels, where questions such as the following
can be asked: ``which properties of the underlying algebraic, update, and
filter parameters are required for closure under a certain operation to
hold?''

\section{Definitions and preliminaries}

A \emph{monoid} is a set equipped with a binary associative operation having
an identity element.  For $(M, +)$ and $(N, \times)$ two monoids, a morphism
$h \colon M \to N$ is a function such that $h(1) = 1$ and $h(ab) = h(a)h(b)$,
for $a, b \in M$.  A \emph{semiring} is a set $K$ equipped with two binary
operations $+$ and $\times$ such that $(R, +)$ is a commutative monoid with
identity $0$, $(R, \times)$ is a monoid with identity $1$ and absorbing
element $0$, and $\times$ distributes over $+$.

A \emph{semiautomaton} $A$ is a tuple $(Q, \Sigma, \delta)$, where $Q$ is a
set of states, $\Sigma$ is an alphabet, and
$\delta \subseteq Q \times \Sigma \times Q$ is a set of transitions.  For a
word $w$, we write $\paths(A, w) \subseteq \delta^*$ for the set of paths on
$A$ labeled $w$.  For a path $\pi$, we write $\orig(\pi)$, \resp
$\dest(\pi)$, for the state in which the path starts, \resp ends.  The
semiautomaton is said to be \emph{unambiguous} if there does not exist two
different paths with same origin, destination, and label.

A \emph{weighted automaton} $W$ over a semiring $K = (K, +, \times)$ is a
tuple $(A, I, U, C)$ such that $A = (Q, \Sigma, \delta)$ is a semiautomaton,
$U\colon \delta^* \to (K, \times)$ is a morphism, and $I, C \colon Q \to K$.
For a word $w \in \Sigma^*$, the automaton computes the value $W(w)$ defined
by:
$$W(w) = \sum_{\pi \in \paths(A, w)} I(\orig(\pi)) \times U(\pi) \times
C(\dest(\pi))\enspace,$$
where $\sum$ is the iterated $+$.  The class of functions computed by
weighted automata on $K$ is denoted $\WA(K)$.  Moreover, $W$ is
\emph{unambiguous} if $A$ is, and we denote $\UWA(K)$ the class of functions
computed by unambiguous weighted automata on $K$---note that in such
automata, the actual interpretation of $+$ is not needed.

\section{Value automata with filters}
Informally, this model differs from a weighted automaton on a semiring
$(K, +, \times)$ in two aspects:
\begin{compactenum}
\item The weighted automaton updates \emph{its only} register $x$ with
  actions of the form $x \leftarrow x \times c$, where $c$ is determined by
  the transition.  In value automata, more than one register can appear, and
  the update expressions can be algebraic expressions.
\item If a word is recognized by $n$ different paths, the values
  $x^{(1)}, x^{(2)}, \ldots, x^{(n)}$ of $x$ at the end of each path are
  \emph{aggregated} using $+$.  In contrast, value automata with filters apply
  a sieve on the different values of $x$, aggregating only those belonging to
  a prescribed set.
\end{compactenum}

With generalization in mind, we wish to present a formalization that does not
impose any \emph{a~priori} structure on the objects at hand, namely the
underlying algebraic structure, the update functions, and the filter sets.

\subsection{Algebraic structures and classes}

\begin{definition}
  An \emph{algebraic structure} $K$ is composed of a base set (written also
  $K$) and multiple internal operations, one of which is a distinguished
  operation called the \emph{aggregate}.  We assume that this operation is
  commutative, can take an unbounded number of arguments (associativity
  implying such a property), and acts as the identity when applied on a
  single element.  This very general concept allows $K$ to be a monoid, a
  semiring, etc.
\end{definition}
  
Let $K$ be an algebraic structure and $f\colon K^m \to K^n$ be a function.
We will always assume that the functions can be expressed using algebraic
expressions, that is: each entry of $f(\vec{x})$ can be defined using an
algebraic expression relying on the operations of $K$ and the variables in
$\vec{x}$.  This corresponds to the idea that $f$ should be an update
function, and should not, for instance, make tests if $K$ is not itself
equipped with a test mechanism.  Further, we say that $f$ is \emph{copyless}
if the set of expressions used to define all the entries of $f(\vec{x})$
contains at most one occurrence of each component of $\vec{x}$.  We say that
$f$ is \emph{moveless} if for all $i$, the $i$-th component of $f(\vec{x})$
is not influenced by the values of $x_j$, $j \neq i$.  Finally, we say that
$f$ is \emph{resetless} if no component of $f(\vec{x})$ is constant when
$\vec{x}$ varies.

\begin{definition}
  An \emph{update class} $\upd$ on $K$ is a function that maps each pair
  $(m, n) \in \bbn \times \bbn$ to a set of functions from $K^m$ to $K^n$.
  We add in superscript \emph{cl}, \emph{ml}, or \emph{rl} when we consider
  the copyless, moveless, or resetless subsets of $\upd$ (e.g., $\upd\Rcm$).
  In our presentation, update classes will serve as restrictions on the
  register updates.
\end{definition}

\begin{examples}
  \begin{compactitem}
  \item When no restriction is imposed, that is, when updates are of the form
    of any algebraic expressions on $K$, we denote the update class $\top$.
  \item Let $+$ and $\times$ be two distinguished operations of $K$.  We let
    \affine be the update class of affine functions, i.e., every
    $f \in \affine(m, n)$ is defined using a matrix $K^{n \times m}$ and a
    vector $\vec{v} \in K^n$ by $f(\vec{x}) = M.\vec{x} + \vec{v}$.  This
    class is succinctly written ``$+, \timesc$'' in~\cite{alur-reg13}, as it
    corresponds to updates using additions and multiplications by constants.
  \item With $K$ as previously, $0$ and $1$ respectively absorbing and
    neutral elements for $\times$, and $m=n$, restrain $\affine$ by having
    $M$ be a $0$--$1$-matrix with exactly one 1 per row.  We obtain the
    update class $\trans$, so denoted as updates are of the form
    $x_i \leftarrow x_j + c$.  This class is written ``\plusc''
    in~\cite{alur-reg13}.
  \item Similarly, when restraining \affine by having $\vec{v}$ be the null
    vector, and $M$ have at most one nonzero entry per row, we obtain the
    update class $\scale$, corresponding to updates of the form
    $x_i \leftarrow x_j \times c$.
  \end{compactitem}
\end{examples}

\begin{definition}
  A \emph{filter class} \filter on $K$ is a map $d \in \bbn \mapsto 2^{K^d}$.
\end{definition}

\begin{examples}
  \begin{compactitem}
  \item The filter class $\top$ maps $d$ to $K^d$.  As a filter, its action
    is then void.
  \item Let $+$ be a distinguished operation of $K$ and $<$ an order on $K$.
    The $\foplus$-definable sets are those expressible as a first-order formula
    using $+$, $<$, and constants.  More precisely, $F \subseteq K^d$ is
    $\foplus$-definable if there is such a formula $\phi$ with $d$ free
    variables such that $\vec{x} \in F$ iff $K \models \phi(\vec{x})$.  We
    write this filter class $\foplus$.  For $K = \bbn, \bbz$ this corresponds to
    the \emph{semilinear sets}, that is, finite unions of sets of the form
    $\vec{v} + K.\vec{v'} + K.\vec{v''} + \cdots$, where the $\vec{v}$'s are in
    $K^d$~\cite{ginsburg-spanier66b}.  We write $\qfplus$ for the
    \emph{quantifier-free} variant of this filter class, that is, sets
    corresponding to formulas with no quantifiers.
  \end{compactitem}
\end{examples}

\subsection{Value automata with filters}

\begin{definition}[Value Automaton with Filter]
  Let $K$ be an algebraic structure, \upd an update class on $K$, and
  \filt a filter class on $K$.  Write $\oplus$ for the aggregate operation
  of $K$.

  A Value Automaton with Filter (VAF) of dimension $d$ is a tuple
  $\calA = (A, I, U, F, C)$ where:
  \begin{compactitem}
  \item $A = (Q, \Sigma, \delta)$ is a semiautomaton,
  \item $I\colon Q \to K^d$ is a partial function called the
    \emph{initialization},
  \item $U\colon \delta \to \upd(d, d)$ is the \emph{update function},
  \item $F \subseteq K^d$ is a subset in $\filt(d)$ called the \emph{filter}.
  \item $C\colon Q \to \upd(d, 1)$ is a partial function called the
    \emph{collapse}.
  \end{compactitem}

  The VAF $\calA$ defines a \emph{partial} function $f\colon \Sigma^* \to K$
  as follows.  Intuitively, for a path in $A$ labeled $w$, the registers are
  initialized with $I(q)$, for $q$ the first state of the path.  They are
  then updated using the functions of $U$, filtered by $F$, and merged into a
  single value by $C(q')$, for $q'$ the last state of the path.  All such
  values for a label $w$ are then aggregated using the aggregate operation.

  Formally, write $U_t$ for $U(t)$, and $C_q$ for $C(q)$.  Define the
  valuation of a path $\val\colon \delta^+ \to K^d$ by
  $\val(t)= I(\orig(t))$, for $t \in \delta$, and
  $\val(\pi t) = U_t(\val(\pi))$, with $\pi \in \delta^*$, $t \in \delta$.
  In particular, if a path starts in $q$ and $I(q)$ is undefined, then the
  valuation of the path is undefined.  Finally:
  $$f(w) = \bigoplus_{\substack{\pi \in \paths(A, w)\\ \val(\pi) \in F}}
  C_{\dest(\pi)}(\val(\pi))\enspace,$$
  where we implicitly discard undefined values of $\val(\pi)$, and the
  $\oplus$ of zero elements is undefined.

  The VAF is deterministic (DetVAF) if $A$ is deterministic and $I$ is
  defined on a single state.  It is unambiguous (UnVAF) if $A$ is.  Finally,
  it is one-success (OneVAF) if
  $|\{\pi \in \paths(A, w) \mid \val(\pi) \in F\}| \leq 1$, that is, if the
  aggregate is not needed.

  The class of such automata is written $\VAF(K, \upd, \filt)$, and similarly
  for DetVAF, UnVAF, and OneVAF.  We identify these classes with the classes
  of functions they define.  For a distinguished element $0$ of $K$, the
  \emph{0-support} of a VAF is set of words it maps to $0$.
\end{definition}

\subsection{Models arising as special cases}

\paragraph{Cost register automata and weighted automata.}  Deterministic VAF
with $\top$ as filter class are identical to the \emph{cost register
  automata} of~\cite{alur-reg13}.  The expressiveness results therein relating
different restrictions of cost register automata are summed up next:
\begin{theorem}[\cite{alur-reg13}]\label{thm:alur}
  With $(K, +, \times)$ a semiring:
  \begin{compactenum}
  \item $\DVAF(K, \scale\Rc, \top)$
  \item[\qquad$\subsetneq$] $\DVAF(K, \top\Rc, \top) = \DVAF(K,
    \scale, \top) = \UWA(K)$
  \item[\qquad$\subsetneq$] $\DVAF(K, \top, \top)$,
  \item $\DVAF(K, \affine\Rc, \top) \subsetneq \DVAF(K, \affine, \top) =
  \WA(K).$
  \end{compactenum}
\end{theorem}

Moreover, weighted automata are naturally expressed as nondeterministic VAF
where only linear transformations are used---in which case, if no register
moves are allowed and no filtering is made, having multiple registers does
not increase the computing power.  Hence, in light of the last point of
Theorem~\ref{thm:alur}:
\begin{proposition}
  With $(K, +, \times)$ a semiring, $\VAF(K, \scale\Rmr, \top) = \DVAF(K,
  \affine, \top)$.
\end{proposition}

Weighted automata over \emph{valuation monoids} constitute a recent fruitful
effort towards generalizing weighted automata to less restricted
settings~\cite{droste_weighted_2011}.  The direction taken there is to move
away from the iterative aspect of weighted automata.  Indeed, a valuation
monoid $M$ is equipped with an extra function $M^+ \to M$ intended to compute
a value given the weights of each of the transitions in a path.  Locality and
incrementality being at the heart of our models, such automata do not seem to
have an equivalent expression within VAF.

\paragraph{Parikh automata and affine Parikh automata.}

Parikh automata were introduced by \kar~\cite{klaedtke-ruess03} and further
studied and extended in~\cite{cafimc11b}.  They are defined as
\emph{recognizers}.  In a Parikh automaton, a set of (integer) counters is
incremented during a run, and a word is accepted if the values belong to a
prescribed semilinear set.  An \emph{affine Parikh automaton} is defined
similarly, except that the update function is lifted to any affine
transformation.  Viewing languages as functions from $\Sigma^*$ to
$\{0, 1\}$, it is readily seen that:

\begin{proposition}
  Deterministic, unambiguous, and nondeterministic Parikh automata (resp.\
  affine Parikh automata) can be simulated by the corresponding sort of
  $\VAF(\bbn, \trans, \foplus)$ (resp.\ $\VAF(\bbn, \affine, \foplus)$).
\end{proposition}

\begin{theorem}[\cite{cafimc11b,cadilhac_unambiguous_2013}]\label{thm:pkh}
  \begin{compactitem}
  \item $\DVAF(\bbn, \trans\Rm, \foplus)$
  \item[\qquad$\subsetneq$] $\UVAF(\bbn, \trans\Rm, \foplus) = \DVAF(\bbn,
    \trans, \foplus) = \UVAF(\bbn, \trans, \foplus)$
  \item[\qquad$\subsetneq$] $\OVAF(\bbn, \trans\Rm, \foplus) \subseteq \VAF(\bbn,
    \trans\Rm, \foplus) = \VAF(\bbn, \trans, \foplus)$
  \item  $\DVAF(\bbn, \affine, \foplus) = \UVAF(\bbn, \affine, \foplus)
    \subsetneq \VAF(\bbn, \affine, \foplus)$,
  \item $\VAF(\bbn, \trans, \foplus)$ and $\DVAF(\bbn, \affine, \foplus)$
    are incomparable.
  \item In these classes, substituting $\bbn$ by $\bbz$ or $\mathbb{Q}$ does
    not impact the the 0-support languages thus defined.
  \end{compactitem}
\end{theorem}

Finally, VAFs can keep an extra register to 1 and use the collapse
function to return it; if the aggregate function of the algebraic structure is
the sum, this counts the accepting paths:
\begin{proposition}
  For any VAF, the function mapping a word to the number of paths whose
  valuations are in the filter is a function in the same class of VAFs.  In
  particular, a VAF can compute the number of accepting paths in a Parikh
  automaton or an affine Parikh automaton.
\end{proposition}


\section{Conclusion}

In this short exposition, we presented a new model that aims at identifying
where some properties of models widely studied in the literature come from.
In a longer version, we will present the closure properties, decidability
properties, and expressiveness results that hold for the general setting.
Our goal is then to identify the essential characteristics that falsify a
given property.  For instance, unambiguity adds no expressiveness as long as
the set of functions $x_i \leftarrow x_j$ is available as updates (as
witnessed here by Theorem~\ref{thm:pkh}).  Similarly, some separations
between the deterministic and nondeterministic variants can be deduced from
the outset, generalizing results for special cases.  Another line of results
is to work towards simplifying the filter set---for instance, using $\qfplus$
instead of $\foplus$ does not impact the expressiveness.  Finally, for
well-behaved algebraic structures, complexity results can be derived for the
decidable problems.

\bibliographystyle{plain}
\bibliography{bib}

\end{document}